# The origin of fermion families and the value of the fine structure constant

J. Lemmon
*154 Cumberland Gap Road*
*Nederland, CO 80466 USA*
*Email: johnjlemmon@gmail.com*

This paper is concerned with a way of thinking about the standard model that explains the existence of three fermion families and the value of the fine structure constant. The main idea is that the ultraviolet divergences that we encounter in the quantum field theories of the standard model, when interpreted appropriately, have a deep physical significance that leads to new relationships between the physical and bare masses of quarks and leptons. This interpretation is based on the assumption of a quantum gravity induced ultraviolet cutoff at the Planck scale and a novel approach to mass renormalization in which the usual perturbation series for the self-mass of a quark or lepton is rearranged and formally summed. Perturbing around the formally summed expression leads to self-consistency equations for the physical quark and lepton masses with multiple solutions that lie outside the reach of conventional perturbation theory. When applied to the standard model at the lowest level of approximation, this approach explains how three generations of charged leptons with a mass spectrum and a value of the fine structure constant in rough agreement with experiment can emerge from a universal bare mass and bare electromagnetic charge. This approach also explains how three generations of physical quark doublets (six flavors) can emerge from a universal bare doublet and bare color charge. Finally, it explains the origin of quark mixing (and CP violation) and the absence of flavor-changing neutral currents.

## 1   Where do the parameters of the standard model come from?

Particle physics is facing a dilemma. The standard model correctly describes the gauge interactions of quarks and leptons, treated as point-like objects with no internal structure. This suggests that the standard model be viewed as a fundamental theory of nature. The problem is that the beauty and simplicity of the standard model is marred by the presence of numerous parameters that appear to defy any deeper explanation and are determined by experiment: multiple generations of quarks and leptons, mass hierarchies, mixing matrices, etc. It seems unthinkable that a theory could be both fundamental and ugly. It is therefore widely believed that the standard model is not truly fundamental, even though we have no evidence of any deeper substructure of quarks and leptons.

The prevailing view is that the standard model is an effective low-energy description of a larger theory based on string theory, supersymmetry, extra dimensions, or some other physics beyond the standard model, despite a lack of experimental evidence. The problem has become more acute now that the first set of runs at the LHC has discovered what appears to be the standard model Higgs boson but not a hint of supersymmetry, extra dimensions, or any other physics beyond the standard model.

We suggest resolving this difficulty from a different point of view: that the multiple generations and associated mass hierarchies and mixings of the fermions are dynamical consequences of a bare Lagrangian which is simple and elegant, and which does not contain multiple generations and mixing. Rather than treating the masses and mixings as incalculable constants, we suggest viewing them as the solutions of self-consistency equations developed from the dynamics of the standard model (with the caveat that some extension of the standard model is necessary to account for neutrino masses and mixing). In this way, we can retain the standard model as both a *description* of what we observe in terms of the physical parameters and an *explanation* of what we observe in terms of a small number of bare parameters.

## 2    The physical significance of ultraviolet divergences

This approach requires an understanding of the relationships between the bare and physical parameters. These relationships (the mass and charge renormalizations) are divergent; the divergences arise because the quarks and leptons are treated as point-like objects, consistent with observation. We therefore expect the divergences to play a central role in any dynamical explanation of mass and charge.

Taking this idea a step further, we propose to work directly with the divergences using a regularization scheme with a momentum-space cutoff $\Lambda$. This amounts to using Pauli-Villars regularization [1] with a single regulator mass. The cutoff is generally viewed as having no physical significance, but simply as a way of defining the mathematics. Here we treat the cutoff as a fundamental scale in nature (related to quantum gravity and the Planck scale) at which the interactions are effectively cut off. This is an old idea dating back to the early days of quantum field theory when the ultraviolet divergences were first discovered [2]. Nowadays, all the various approaches to quantum gravity (string theory, loop quantum gravity, noncommutative geometry, doubly special relativity, etc.), in one way or another, invoke the notion of a minimal length (or equivalently, a high-energy cutoff) [3]. What is different about the present approach is that rather than simply *assuming* that the cutoff is on the order of the Planck scale, we shall *determine* a value of the cutoff from the masses of the charged leptons, whose values *require* a cutoff on the order of the Planck scale.

Promoting the role of the cutoff to that of a physical observable brings with it the well-known problem of ambiguities since regularization of the divergent integrals with a momentum-space cutoff is not a unique procedure. Here we view the cutoff as a consequence of quantum gravity, for which we currently have no operational theory. Such a theory would presumably enable one to describe the short-distance structure of space-time and its effects on the fundamental interactions in a way that leads to unambiguous, finite and gauge invariant results. In the meantime, we shall attribute these difficulties to the lack of a consistent theory of quantum gravity and view this procedure for dealing with the divergences as an *effective description* of the effects of quantum gravity on mass and charge renormalization. Specifically, we propose to deal with the divergences as follows.

As stated above, we use Pauli-Villars regularization with a single cutoff mass. This regularization technique leads to gauge invariant results in quantum electrodynamics, but produces gauge-dependent mass renormalizations in the electroweak theory, due to parity violation in the weak interactions. However, it has been shown that electrodynamics can be formulated in a gauge-free framework [4], in which gauge ambiguities are absent from the outset. The theory is formulated with a physical vector potential that does not involve unphysical (gauge) degrees of freedom. Quantizing the theory leads to a photon propagator that is gauge invariant by construction. It is noteworthy, as pointed out by these authors, that although the physical photon field is transverse, the resulting photon propagator does not correspond to what is the Landau (transverse) gauge in the usual formulation but corresponds to the Feynman gauge. The gauge-free formulation has been applied to the Coleman-Weinberg mechanism [5] of mass generation via radiative corrections, which also suffers from a lack of gauge invariance [6] due to the gauge dependence of the effective potential, thereby eliminating the gauge dependence of this mechanism and rendering the masses to be physical.

We shall adopt a similar strategy here and assume that the Feynman gauge corresponds to physical gauge-free propagators not only of the photon, but also of the weak bosons in the electroweak theory. Although we have no proof of this, in the usual ($R_\xi$) formulation the masses of the Goldstone bosons that correspond to the longitudinal degrees of freedom of the massive gauge bosons are gauge-dependent; the Feynman gauge is that gauge in which these masses are equal to the masses of the corresponding gauge bosons. Since the Goldstone bosons correspond to physical degrees of freedom of the gauge bosons, it seems natural that they should share the same mass, again suggesting the use of the Feynman gauge to calculate physical masses.

Even after specifying the regularization procedure as described above, there remain small (cutoff-independent) contributions to the mass renormalizations that are ambiguous, depending on which mass (fermion or boson) is replaced by the cutoff when regularizing the fermion self-energy. However, it turns out that these relatively small contributions have an insignificant effect on the numerical results presented in this paper. We shall adopt the procedure whereby the boson mass is replaced by the cutoff in the fermion self-energy. In the vacuum polarization, only the fermion mass appears in the (one-loop) contribution, and therefore the cutoff replaces its mass. This procedure leads to the well-known lowest-order expressions for the mass and charge renormalization of the electron in quantum electrodynamics, originally written down by Feynman [7] over sixty years ago.

## 3  The generation problem

This work has been motivated by an attempt to understand the existence of three generations of quarks and leptons using only the *known laws of physics*, without introducing extraneous particles, interactions, or symmetries that have no observational support. We begin by considering the mass of the electron in quantum electrodynamics.

The mass of the electron arises from the bare mass and the radiative corrections to the bare electron propagator,

$$S(p) = \frac{i}{p \cdot \gamma - m_0}, \tag{1}$$

where $m_0$ is the bare mass, $p$ is the electron four-momentum, and $\gamma$ is a Dirac matrix. Denoting the electron proper self-energy by $\Sigma(p)$, the dressed propagator is

$$S'(p) = S(p) + S(p)\big(-i\Sigma(p)\big)S(p) + S(p)\big(-i\Sigma(p)\big)S(p)\big(-i\Sigma(p)\big)S(p) + \cdots, \tag{2}$$

which can be formally summed as a geometric series, so that

$$S'(p) = \frac{i}{p \cdot \gamma - m_0 - \Sigma(p)}. \tag{3}$$

The physical (pole) mass is defined as the location of the pole in the propagator, and is therefore the sum of the bare mass and the self-mass, defined as the proper self-energy $\Sigma(p)$ evaluated on the mass-shell, $p \cdot \gamma = m$, where $m$ is the physical mass. On dimensional grounds the self-mass is proportional to $m$, and is therefore the product of $m$ and a dimensionless quantity $\delta(m)$:

$$m = m_0 + m \cdot \delta(m). \tag{4}$$

If $\delta(m)$ were finite we could solve (4) self-consistently for $m$; however, $\delta(m)$ is formally infinite and in the usual approach to mass renormalization we add a counterterm $(m_0 - m)$ to the Dirac equation and treat the counterterm as an additional interaction, cancelling the self-mass wherever it arises in the calculation of scattering amplitudes. This amounts to simply ignoring the self-mass at every order of perturbation theory and is perfectly fine for calculating scattering amplitudes in terms of the physical mass. The problem is that while this enables us to avoid dealing directly with the divergences, it precludes any possibility of understanding the relationship between the bare and physical mass (even if the self-mass were finite), and hence, of gaining any insight into the fact that in nature there are three different physical masses of the electron (charged leptons).

On the other hand, treating the cutoff as a physical observable leads to a self-consistency equation for the electron mass that allows us to explore the relationship between the bare and physical masses. We are seeking multiple solutions that can be identified with the masses of the charged leptons: the presence of multiple generations can then be understood as a dynamical consequence of a simpler bare theory in which multiple generations are absent.

Using the regularization procedure described in the previous section, the lowest order (one loop) perturbative expression for $\delta(m)$ in quantum electrodynamics is [7]

$$\delta^{(1)}(m) = \frac{3\alpha_0}{4\pi}\left(\ln\frac{\Lambda^2}{m^2} + \frac{1}{2}\right), \tag{5}$$

where $\alpha_0$ is the bare coupling constant. Approximating $\delta(m)$ by $\delta^{(1)}(m)$, substituting (5) into (4) and solving for $m$, it is easy to see that for negative values of the bare mass there can be zero, one, or two solutions for the physical mass, depending on the value of the bare mass, and for positive values of the bare mass there is always one solution for the physical mass. This can be seen from inspection of graphs of the functions $m$ and $m_0 + m\delta^{(1)}(m)$ vs. $m$, as shown in Figure 1. Values of $m$ where $m_0 + m\delta^{(1)}(m) = m$ correspond to values of the physical mass.

Note that a negative bare mass is perfectly acceptable if the physical masses are positive. In the standard model masses are generated via the Higgs mechanism, in which fermion masses in the Lagrangian are of the form $Gv$, where $G$ is a Yukawa coupling constant and $v$ is the vacuum expectation value of the Higgs field (246 GeV). If we interpret the bare mass $m_0$ as $G_0 v$, where $G_0$ is a bare Yukawa coupling constant, a negative bare mass simply corresponds to a negative value of $G_0$. Since the physical mass is the sum of the bare mass and the self-mass, a physical Yukawa coupling $G$ can be defined by

$$m = Gv = G_0 v + m\delta(m), \tag{6}$$

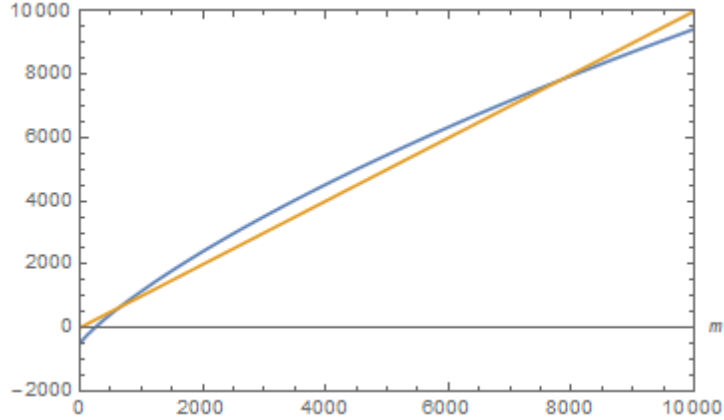

Figure 1. Plots of the functions $m$ and $m_0 + m \cdot \delta^{(1)}(m)$ vs. $m$ (in arbitrary units) with a negative value of $m_0$. Values of $m$ where the functions intersect correspond to values of the physical mass.

so that the physical mass in the Lagrangian retains its customary form and is positive. However, the fact remains that, regardless of the values of the bare parameters, lowest-order perturbation theory can account for no more than two physical masses. If higher-order corrections were to alter this conclusion, then perturbation theory would be unreliable.

## 4  Self-mass as a formal sum

Perturbation theory is unable to explain the existence of three generations of charged leptons. This is because the perturbation series for the self-mass arises from an *infinite* sequence of radiative corrections to the bare electron propagator. In diagrammatic language, the one-loop correction generates a two-loop correction, which generates a three-loop correction, ad infinitum. Every time we include a new radiative correction in a diagram, new electron propagators are introduced in the internal lines in the diagram, which then introduce more radiative corrections, etc.

The problem is that it's logically inconsistent to use perturbation theory to develop a self-consistency equation for the mass of the electron that only takes a *finite* number of radiative corrections into account when perturbation theory tells us that the electron mass comprises an *infinite* sequence of radiative corrections. This inconsistency can be avoided by treating the self-mass of the electron as a formal sum of an infinite sequence that automatically takes into account contributions from *all* orders of perturbation theory. Only in this way can one account for radiative corrections creating radiative corrections *ad infinitum*. We will show that the difficulty explaining the origin of three generations is not due to the limitations of perturbation theory per se but is due to a failure to restructure the perturbation series into a form that can be formally summed.

This formal summation of a perturbation series is not necessary in the calculation of scattering amplitudes because there one is developing *expressions* that are evaluated numerically (cross sections, anomalous magnetic moments, Lamb shift, etc.). If the expansion parameter is small, one can terminate the series and obtain numerically accurate results, as we ordinarily do in quantum electrodynamics. Here, however, we are attempting to develop a self-consistency *equation* for the electron mass, and the functional dependence of the self-mass on the physical mass is of critical importance.

Using the same rules over and over to build up higher loop diagrams and calculate their contributions to the self-mass suggests rearranging the perturbation series for the self-mass using a simple multiplicative rule that is applied repetitively. The self-mass is proportional to $m$, so the simplest way to generate such a sequence is to multiply each term in the sequence by the same dimensionless function $\Delta(m)$, where $m\Delta(m)$ is the first term, to obtain the next term. We shall therefore write the self-mass as $m\Delta(m)[1 + \Delta(m) + \Delta^2(m) + \Delta^3(m) + \cdots]$. This expression can be formally summed in analogy to the way the self-energy corrections to the bare electron propagator in (2) are formally summed to obtain the simple closed-form expression for the dressed propagator in (3). Thus, we propose replacing (4) with

$$m = m_0 + m\frac{\Delta(m)}{1 - \Delta(m)}. \tag{7}$$

The function $\Delta(m)$ can be developed iteratively from $\delta(m)$ by expanding $\Delta(m)$ in a power series in $\alpha_0$ in such a way that $\frac{\Delta(m)}{1-\Delta(m)}$ is formally equivalent to the perturbation series for $\delta(m)$. If one writes the perturbation series for $\delta(m)$ as $\delta(m) = \delta^{(1)}(m) + \delta^{(2)}(m) + \delta^{(3)}(m) + \cdots$ and the power series expansion of $\Delta(m)$ as $\Delta(m) = \Delta^{(1)}(m) + \Delta^{(2)}(m) + \Delta^{(3)}(m) + \cdots$ (where the superscripts denote powers of $\alpha_0$), then setting $\delta(m) = \frac{\Delta(m)}{1-\Delta(m)}$ and equating equal powers of $\alpha_0$ leads to $\Delta^{(1)}(m) = \delta^{(1)}(m)$, $\Delta^{(2)}(m) = \delta^{(2)}(m) - \delta^{(1)^2}(m)$, etc. In this way, the n[th] term in the expansion of $\Delta(m)$ can be determined from the first n terms in the expansion of $\delta(m)$. If one calculates perturbatively in the usual way up to the n[th] order, in effect one is setting all the higher order terms to zero. After n iterations of the present scheme, one has developed a form for the structure in (7) that agrees with perturbation theory up to the n[th] order, but which also contains what can be viewed as estimates of all the higher order terms. Rather than using perturbation theory to directly calculate the self-mass $\delta(m)$, in which case the sequence of self-mass corrections is inevitably terminated at some point and therefore fails to describe the characteristic that radiative corrections create radiative corrections ad infinitum, we use perturbation theory to develop the rule $\Delta(m)$ by which radiative corrections create radiative corrections.

There remains a question of the convergence of the solutions of (7) in higher orders, for which we have no proof. However, the dominant contributions to the self-mass come from the cutoff-dependent contributions, which originate from large momenta running around the loops. Since the cutoff-independent contribution to the photon propagator tends to cancel the cutoff-dependent photon wave function renormalization at high energies [8] (at least in lowest order), the photon propagator corrections are not expected to make dominant contributions to the self-mass. The cutoff-dependent contributions from the vertex and electron wave function renormalizations cancel via the Ward identity. The only remaining cutoff-dependent contributions come from the mass renormalizations. At each successive order of perturbation theory these arise from the insertion of a *lowest-order self-energy correction* into an electron propagator, which generates another factor of $\delta^{(1)}(m)$. This suggests that the dominant contributions to the self-mass may behave like powers of $\delta^{(1)}(m)$, and hence, of $\Delta^{(1)}(m)$. This in turn suggests that the lowest order expression for $\Delta(m)$ may approximately describe the mass spectrum when substituted into (7). In fact, as shown below, when one uses perturbation theory in this way with the usual Feynman rules, the one-loop expressions provide a description of the charged lepton mass spectrum and a value of the fine structure constant in qualitative agreement with experiment. The formal summation is of primary importance for understanding the mass spectrum.

## 5 Spinor electrodynamics – a simple example

In this section we apply these ideas to the quantum electrodynamics of electrons and photons. Approximating $\Delta(m)$ by $\Delta^{(1)}(m) = \delta^{(1)}(m)$ and substituting (5) into (7) leads to a structure of the self-mass entirely different from that using (4), as can be seen in Figure 2, where we have plotted the functions $m_0 + m \frac{\Delta^{(1)}(m)}{1-\Delta^{(1)}(m)}$ and $m$ vs. $m$. There can be either one or three solutions for the physical mass, depending on the values of the bare mass and charge. In Figure 2 we have chosen values that give three solutions and have plotted the functions with three different scales to clearly reveal all three solutions (which exhibit a large hierarchy) as well as the singularity where $\Delta^{(1)} = 1$. This large hierarchy is generated by a small coupling constant ($(\frac{\alpha_0}{\pi}) \approx 0.04$), reflecting the nonperturbative nature of the solutions. We interpret these three solutions as the masses of the electron ($m_e$), the muon ($m_\mu$), and the tau lepton ($m_\tau$). All three solutions lie in the region where $\Delta^{(1)}(m) < 1$, and therefore within the radius of convergence of the formally summed geometric series. (In the region where $\Delta^{(1)}(m) > 1$ the series diverges and we treat the self-mass as infinite.)

### 5.1 Determination of the bare parameters and the cutoff

Equations (5) and (7) are unable to simultaneously account for the observed masses of all three charged leptons for any values of the bare parameters and the cutoff. We shall therefore use values that correctly determine two of the masses, and then use these values to predict the third mass. We can then use these values of the bare coupling constant, the cutoff, and the physical masses to calculate a value of the fine structure constant *from first principles*.

As pointed out below, the fine structure constant depends upon all three physical masses but is least dependent on the heaviest mass (tau lepton). We shall therefore use values of the bare parameters and the cutoff that correctly predict the masses of the electron and muon, and then use these values to predict values of the tau lepton mass and the fine structure constant.

We have found that the parameters

$$\alpha_0 = 0.130; \ m_0 = -103.3 \ MeV; \ \Lambda = 3.65 \ TeV \tag{8}$$

when substituted into (5) and (7) predict the mass spectrum

$$m_e = 0.511 \ MeV; \ m_\mu = 105.7 \ MeV; \ m_\tau = 1.002 \ GeV. \tag{9}$$

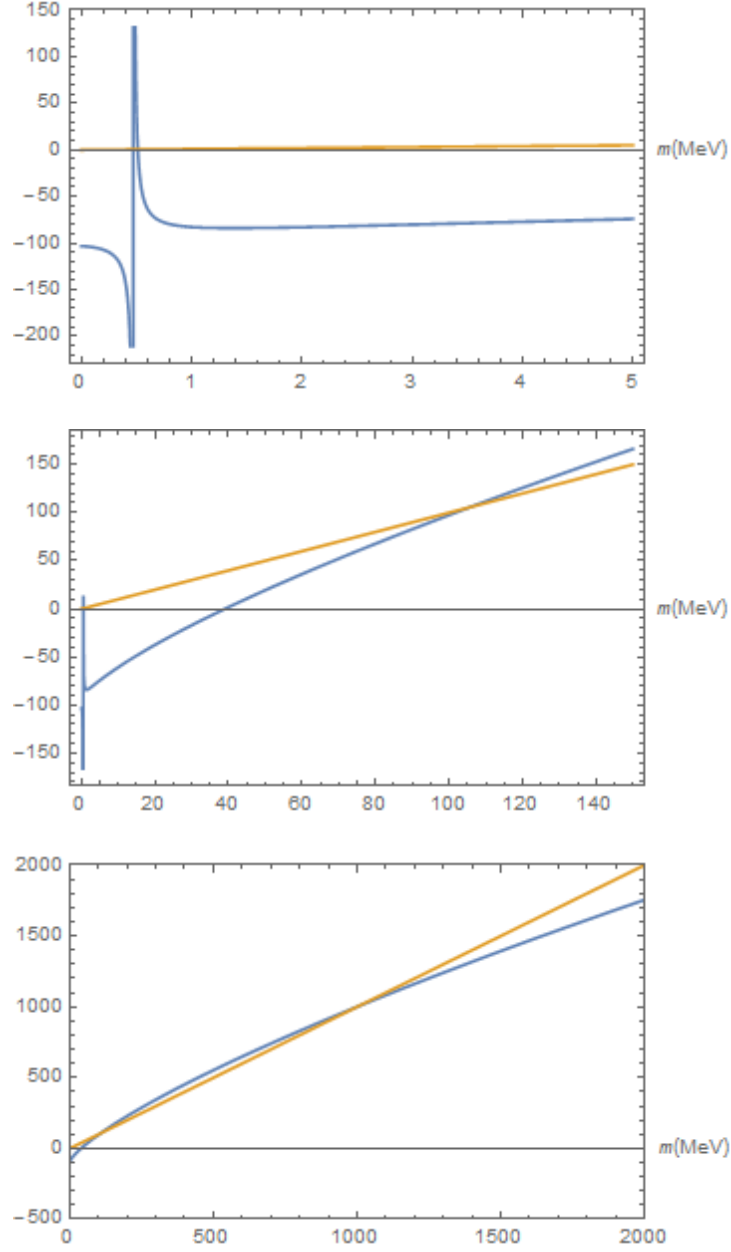

Figure 2. Plots of the functions $m$ and $m_0 + m \frac{\Delta^{(1)}(m)}{1-\Delta^{(1)}(m)}$ vs. $m$ using the values of the bare parameters discussed in the text. Values of $m$ where the functions intersect correspond to the values of the charged lepton masses.

Although this prediction of the tau mass is less than the observed value ($1.777\ GeV$), we view the prediction of a third lepton mass with the correct order of magnitude while simultaneously explaining the masses of the electron and the muon as a qualitative success.

Interpreting the TeV scale cutoff as a fundamental Planck scale is reminiscent of low scale gravity models with large extra dimensions. However, the effect of the weak interactions will be shown in the following section to drive the cutoff up from the TeV scale to the usual Planck scale.

## 5.2 Calculation of the fine structure constant

In this approach to the generation problem, the masses of the charged leptons are the key to understanding their charge. *The existence of three generations of charged leptons with a large mass hierarchy requires a specific value of the bare coupling constant, from which a rough estimate of the fine structure constant is a simple, straightforward exercise*. Remarkably, the weak interactions have little effect on this estimate, as shown in the following section.

The fine structure constant is the renormalized coupling constant, defined by

$$\alpha = Z_3 \alpha_0, \tag{10}$$

where $Z_3$ is the photon wave function renormalization constant. The lowest order contribution to $Z_3$ from the leptons is [7]

$$Z_3 = 1 - \frac{\alpha_0}{3\pi}\left(\ln\frac{\Lambda^2}{m_e^2} + \ln\frac{\Lambda^2}{m_\mu^2} + \ln\frac{\Lambda^2}{m_\tau^2}\right), \tag{11}$$

where we have added the terms corresponding to the muon and the tau lepton, which are not included in [7]. The quantity in parentheses (times $\frac{\alpha_0}{3\pi}$) can be viewed as a polarization charge due to the polarization of the vacuum; larger masses make smaller contributions to the polarization. The cutoff dependence arises because interactions of the leptons and photons are treated as point-like. Note that in contrast to our treatment of the electron mass, here we are not developing a self-consistency equation, but are simply evaluating an expression for the vacuum polarization expressed as a power series in the coupling constant, as we ordinarily do in quantum electrodynamics; the only difference is that here we are expanding in powers of the bare coupling constant rather than the physical coupling. As in our treatment of the electron mass, there remains a question of the convergence of this series.

The contribution to the vacuum polarization from hadrons must be added to this expression. Since the hadrons are extended composite objects, the point-like interactions that produce the cutoff dependence are replaced by electromagnetic form factors $G_E(Q^2)$ and $G_M(Q^2)$ that result in relatively small, cutoff-independent corrections ($Q^2$ is the momentum transfer squared). Physically, these form factors correspond to the momentum space representations of the electric charge and magnetic moment density distributions inside the hadrons. Since hadrons have a spatial extent $\sim 1\ fm \sim (1\ GeV)^{-1}$, we expect the form factors to fall off rapidly when $Q^2 \gg (1\ GeV)^2$. For example, the electromagnetic form factors of the proton are well described up to $10\ GeV^2$ by the so-called dipole function

$$G_E(Q^2) = \frac{G_M(Q^2)}{\mu_P} = \left(1 + \frac{Q^2}{0.71\ GeV^2}\right)^{-2},$$

where $\mu_P$ is the magnetic moment of the proton. At larger values of $Q^2$, $G_M$ falls somewhat faster than $1/Q^4$.

Since the one-loop vacuum polarization diagrams introduce products of two such factors (one at each vertex), we expect the contributions to the polarization charge due to proton loops to be suppressed by factors $\lesssim 10^{-16}$ when $Q^2 \gtrsim 100\ GeV^2$. The contributions from all hadrons, which have various form factors, must be included, but even slower decreases in the form factors (e.g. $1/Q^2$) effectively cut off the vacuum polarization at these values of $Q^2$. We shall therefore neglect the hadronic contributions to the polarization charge when $Q^2 \gtrsim 100\ GeV^2$, and estimate the hadronic polarization charge as the hadronic contribution to the running of the fine structure constant from zero energy up to the Z boson mass (91 GeV).

This running of the fine structure constant can be parameterized as

$$\alpha(M_Z^2) = \frac{\alpha(0)}{1 - \Delta\alpha_L(M_Z^2) - \Delta\alpha_H(M_Z^2)}, \tag{12}$$

where $\Delta\alpha_L$ and $\Delta\alpha_H$ are the leptonic and hadronic contributions, respectively. The most precise value of $\Delta\alpha_H(M_Z^2)$ has been determined [9] to be 0.02757(8). Since $\alpha(M_Z^2) \approx 0.007818 \approx \frac{1}{127.909}$, our estimate of the hadronic contribution to the polarization charge is $\alpha(M_Z^2)\Delta\alpha_H(M_Z^2) = 0.00021554$.

Using the values of the parameters in (8) and the values of the *predicted* physical masses in (9), our one-loop estimate of the fine structure constant is

$$\begin{aligned}\alpha &= \alpha_0\left[1 - \frac{\alpha_0}{3\pi}\left(\ln\frac{\Lambda^2}{m_e^2} + \ln\frac{\Lambda^2}{m_\mu^2} + \ln\frac{\Lambda^2}{m_\tau^2}\right)\right] - \alpha(M_Z^2)\Delta\alpha_H(M_Z^2) \\ &= 0.00630237 \approx \frac{1}{158.7}.\end{aligned} \tag{13}$$

A lowest order calculation using only quantum electrodynamics has led to an estimate of the fine structure constant that agrees with experiment to within 15%. The fact that the measured value of $\alpha \sim \frac{1}{20} \cdot \alpha_0$ indicates that roughly 95% of the bare coupling constant is shielded by the vacuum polarization, which means that our estimate of the polarization charge is accurate to within 1%.

A question that has often been asked is: why is a fundamental dimensionless constant like the fine structure constant so small, and not of order unity? From the present perspective, the fine structure constant is not truly fundamental, but is a derived quantity. The fundamental quantity is the bare charge of the electron, and since $\alpha_0 \equiv \frac{e_0^2}{4\pi} \approx 0.130$, it follows that the bare charge is approximately

$$e_0 \approx 1.28, \tag{14}$$

which is clearly of order unity.

## 6 Extension to the electroweak theory

### 6.1 The use of a single gauge coupling constant

In our treatment of the self-mass of the electron in QED we expressed the self-mass in terms of the *bare* electromagnetic coupling constant and the *physical* electron mass to obtain a self-consistency equation for the mass. We then used the resulting physical masses, the bare coupling constant, and the cutoff to calculate the physical coupling constant. We wish to extend this strategy to the electroweak theory.

The problem is that the spontaneous breakdown of the $SU(2)_L \times U(1)_Y$ gauge symmetry down to the $U(1)$ of electromagnetism results in relationships among the weak boson masses $M_W$ and $M_Z$, the two gauge coupling constants $g$ and $g'$ associated with the $SU(2)_L$ and $U(1)_Y$ gauge groups, respectively, the electromagnetic coupling $e$, and the weak mixing angle $\theta_W$. The mixing angle can be viewed as a function of the coupling constants $g$ and $g'$ (for example, $\frac{g'}{g} = \tan\theta_W$) or as a function of the masses $M_W$ and $M_Z$ (for example, $\frac{M_W}{M_Z} = \cos\theta_W$ in the absence of loop corrections). If we wish to calculate in terms of bare coupling constants and physical masses, we are therefore left with a choice: do we treat the mixing angle as a function of the bare coupling constants or the physical masses? Since we wish to follow as closely as possible the procedure we used in QED, we propose to calculate in terms of the *bare* electromagnetic coupling $e_0$ and the *physical* mixing angle, viewed as a dimensionless function of the *physical* weak boson masses. This amounts to rewriting the electroweak Lagrangian, when expressed in terms of $g$ and $g'$, with the coupling constant replacements

$$\frac{gg'}{\sqrt{g^2 + g'^2}} \to e_0, \tag{15.1}$$

$$g \to e_0 \csc\theta_W, \tag{15.2}$$

$$\frac{g^2}{\sqrt{g^2 + g'^2}} \to e_0 \cot\theta_W, \tag{15.3}$$

$$\frac{g'^2}{\sqrt{g^2 + g'^2}} \to e_0 \tan\theta_W, \tag{15.4}$$

and calculating the mass spectrum and the fine structure constant using the *bare* electromagnetic coupling $e_0$ and the *physical* mixing angle $\theta_W$.

We stress that this is an assumption that we cannot justify a priori; however, the electromagnetic coupling and the mixing angle are the quantities that are directly measured (as opposed to $g$ and $g'$) and this approach seems to be the most natural and simplest extension of our calculations in electrodynamics to the electroweak theory. It may seem inconsistent to calculate in terms of the bare electromagnetic coupling $e_0$ in the right-hand sides of the replacements in (15.1-4) while not replacing $\theta_W$ by its bare value, expressed in terms of $g_0$ and $g'_0$. However, as shown below, the electroweak theory expressed in terms of the bare electromagnetic coupling and the mixing angle generates a mass spectrum and value of the fine structure constant in approximate agreement with experiment when the mixing angle is equal to the *physical* value that is observed. Nature seems to be telling us to view the electroweak theory in terms of a *single* gauge coupling constant (the electromagnetic coupling), which corresponds to a conserved charge and plays a special role in physics. The coupling of quarks and leptons to the weak gauge bosons involves this single gauge coupling constant and dimensionless functions of the boson masses, which can be simply expressed as trigonometric functions of the mixing angle. In this way we can self-consistently calculate the mass spectrum and the fine structure constant in terms of the bare electromagnetic coupling and the physical masses in the theory, in close analogy to our calculations QED.

## 6.2   Mass and charge renormalizations

We have used this strategy and the Feynman rules of the electroweak theory to calculate the lowest order mass and charge renormalizations of the electron. In calculating the mass renormalization a complication arises in the electroweak theory not encountered in electrodynamics, due to the chirality of the weak interactions. The electron self-energy $\Sigma(p)$ can be written in terms of "right- and left-handed" contributions $\Sigma_R(p)$ and $\Sigma_L(p)$ as $\Sigma(p) = \Sigma_R(p)\gamma_R + \Sigma_L(p)\gamma_L$, where $\gamma_{R,L}$ are the usual right- and left-handed helicity projection operators, $\gamma_{R,L} \equiv \tfrac{1}{2}(1 \pm \gamma_5)$. Evaluating $\Sigma_{R,L}$ on the mass shell results in "right- and left-handed" contributions to the self-mass that can be written as $m_R$ and $m_L$, respectively. Due to parity violation in the weak interactions, $m_R \neq m_L$. However, it has been shown [10] that in a parity violating theory, *to lowest order*, the pole mass is given by $m = G_0 v + \tfrac{1}{2}(m_R + m_L)$, where we have written the vacuum mass $G_0 v$ in place of the bare mass $m_0$. We have used this rule in calculating the mass renormalization.

In analogy to the lowest order expression for the mass renormalization in quantum electrodynamics given by (5), we have found

$$\delta^{(1)} = \Delta^{(1)} = \frac{3\alpha_0}{4\pi}\left(\ln\frac{\Lambda^2}{m^2} + \frac{1}{2}\right) - \frac{\alpha_0}{\pi}\left[\frac{1}{16}csc^2\theta_W I_1(m,m_\nu)\right. \\ \left. + \left(\frac{1}{32}cot^2\theta_W + \frac{5}{32}tan^2\theta_W - \frac{1}{16}\right)I_2(m) + \left(\frac{1}{2} - \frac{1}{2}tan^2\theta_W\right)I_3(m)\right], \quad (16)$$

where the Feynman integrals $I_{1,2,3}(m)$ are

$$I_1(m,m_\nu) = 2\int_0^1 z\,\ln\frac{\Lambda^2 z}{M_W^2 z - m^2 z(1-z) + m_\nu^2(1-z)}dz, \quad (17.1)$$

$$I_2(m) = 2\int_0^1 z\,\ln\frac{\Lambda^2 z}{M_Z^2 z + m^2(1-z)^2}dz, \quad (17.2)$$

$$I_3(m) = \int_0^1 \ln\frac{\Lambda^2 z}{M_Z^2 z + m^2(1-z)^2}dz, \quad (17.3)$$

and $m_\nu$ is the neutrino mass (which we are treating as zero in this paper).

The expression in (16) includes contributions from the gauge bosons but does not include contributions from the Goldstone and Higgs bosons, since they are numerically insignificant. This is because the scalar bosons couple to the leptons with the bare Yukawa coupling $G_0$. Since we know from (8) that the bare mass $m_0 = G_0 v = -103.3\ MeV$ and the measured value of $v = 246\ GeV$, it follows that the coupling $\frac{G_0^2}{4\pi} \sim 10^{-8}$, which is insignificant compared to $\alpha_0 \sim 10^{-1}$. (As explained below, these values of the bare parameters are only slightly changed in the presence of the weak interactions.)

Substituting (16) and (17.1-3) into (7) again generates three solutions (with appropriate choices of the bare parameters and the cutoff) that can be identified with the masses of the charged leptons. The effect of the weak interactions, corresponding to the terms in the square brackets in (16), can be understood as follows. Since the lepton masses are much less than the weak boson masses, the integrals in (17.1-3) are approximately equal to $ln\frac{\Lambda^2}{M_{W,Z}^2}$ . It is then easy to show that the upper and lower bounds on the solutions corresponding to $\Delta^{(1)} = ½$ and $\Delta^{(1)} = 1$, respectively, are both (approximately) scaled by the factor

$$\left(\frac{\Lambda}{M_W}\right)^{-\frac{4}{3}\left(\frac{1}{16}csc^2\theta_W\right)} \left(\frac{\Lambda}{M_Z}\right)^{-\frac{4}{3}\left(\frac{1}{32}cot^2\theta_W - \frac{11}{32}tan^2\theta_W + \frac{7}{16}\right)} \tag{18}$$

so that the solutions of the self-consistency equation are also approximately scaled by this factor and their ratios are only slightly changed. Thus, the structure of the self-consistency equation is qualitatively similar to that in QED, except for a change in scale. Since the mass scale is set by the cutoff and the mass ratios are determined by the bare parameters, it follows that the effect of the weak interactions is to rescale the cutoff (by an amount determined by the weak boson masses and the weak mixing angle) and leave the bare parameters only slightly changed.

In order to calculate the lepton masses, we need measured values of the boson masses and the mixing angle to be substituted into (21) and (22.1-3). For the masses we have used

$$M_W = 80.39\ GeV;\ M_Z = 91.19\ GeV. \tag{19}$$

Due to the scale factor in (18), the cutoff (and hence, the mass spectrum and the fine structure constant) are sensitive functions of the mixing angle. The mixing angle runs with energy, and it is therefore important to use a value measured at the energy scale corresponding to the lepton masses ($\lesssim 1.8\ GeV$). (The mixing angle does not vary appreciably over this range of energies.) The most precise measurements in this energy range are an atomic parity violating experiment [11] (corresponding to zero energy) that reported a value of $sin^2\theta_W = 0.2381 \pm 0.0011$ and SLAC Experiment 158 [12] that measured Moller scattering at an energy of $0.16\ GeV$ and found $sin^2\theta_W = 0.2397 \pm 0.0013$. Since the central values differ in the third significant figure but have comparable uncertainties, we have chosen to estimate the mixing angle as their mean value to three significant figures, that is,

$$sin^2\theta_W \approx 0.239, \tag{20}$$

which lies within the reported uncertainties of both experiments.

Turning now to the charge renormalization, the electroweak theory introduces an additional contribution to the vacuum polarization at lowest order from a W boson loop.[1] Again using the Feynman rules for the electroweak theory, we have found, in analogy to the contributions in (11) from the lepton loops, that the W boson loop makes a contribution

---

[1] In principle, there are also contributions from the charged Goldstone boson loops. However, these are cancelled by the contributions from the loops of charged Fadde'ev-Popov ghosts.

$$+\frac{11\alpha_0}{12\pi} ln \frac{\Lambda^2}{M_W^2}$$

to the photon renormalization constant $Z_3$. Note that this contribution is positive, in contrast to the negative contributions from the lepton loops in (11). The lowest order expression for the fine structure constant then becomes

$$\alpha = \alpha_0 \left[1 - \frac{\alpha_0}{3\pi}\left(ln\frac{\Lambda^2}{m_e^2} + ln\frac{\Lambda^2}{m_\mu^2} + ln\frac{\Lambda^2}{m_\tau^2}\right) + \frac{11\alpha_0}{12\pi} ln\frac{\Lambda^2}{M_W^2}\right] - \alpha(M_Z^2)\Delta\alpha_H(M_Z^2). \tag{21}$$

## 6.3   Calculations of the mass spectrum and the fine structure constant

We are now in a position to calculate the masses of the charged leptons and the fine structure constant from the equations of the electroweak theory. We follow the same strategy we used in QED by choosing values of the bare parameters and the cutoff that generate the measured values of the electron and muon masses. Substituting the values of the weak boson masses and the weak mixing angle in (19) and (20) into (18) then *requires* the cutoff to be increased approximately *sixteen orders of magnitude* from the TeV scale to the Planck scale. We shall therefore identify the cutoff with the Planck mass $M_P$, so that

$$\Lambda = M_P = 1.22 \cdot 10^{19} \, GeV, \tag{22}$$

which is consistent with our assumption of a quantum gravity induced minimal length. *The weak interactions are why the lepton masses are so remarkably small compared to the Planck mass*.

Substituting (16) and (17.1-3) into (7) with the bare parameters

$$\alpha_0 = 0.13499; \, m_0 = -91.27 \, MeV, \tag{23}$$

the weak boson masses and mixing angle in (19) and (20), and the cutoff in (22) generates the physical masses

$$m_e = 0.511 \, MeV; \, m_\mu = 105.67 \, MeV; \, m_\tau = 667.7 \, MeV. \tag{24}$$

As we found in QED, bare parameters that generate the measured values of the electron and muon masses predict a value of the tau mass less than the measured value. However, in natural units where $M_P = 1$, the measured value of the tau mass is $10^{-18.84}$ versus our prediction of $10^{-19.26}$, a discrepancy of only 2.2% in orders of magnitude.

The dramatic increase in the value of the cutoff (from the TeV scale to the Planck scale) potentially threatens to spoil the approximate agreement between the measured and predicted values of the fine structure constant that we found in QED. However, substituting the values of the cutoff, the bare coupling constant, the physical masses in (24), and the hadronic correction into (21), we find that

$$\alpha = 0.00607757 \approx \frac{1}{164.5}. \tag{25}$$

Thus, the W and Z boson contributions to the mass renormalization that drive up the cutoff are compensated by the W boson contribution to the charge renormalization to maintain the delicate balance between the bare coupling constant $\alpha_0$ and the shielding due to vacuum polarization ($\approx -0.95\alpha_0$). We view this as strong support for our assumptions, since even small changes in the peculiar coefficients in the mass and charge renormalizations in (16) and (21) would upset this approximate agreement between theory and experiment.

## 7 Quark masses and mixing

### 7.1 Quark mass renormalizations

The idea of restructuring the perturbation series for the self-mass of a fermion as a formal sum can be applied to quarks as well as leptons. We have derived lowest-order expressions for the quark self-masses that are analogous to (16). The lowest-order expressions for the quark self-energies in QCD have the same structure as those in QED and fail to take into account the non-Abelian character of QCD. Consequently, we do not necessarily expect the one-loop approximation to predict realistic quark masses, particularly for the light quarks, since QCD becomes nonperturbative at low energies. Our intention has been to see if the one-loop approximation can at least explain why quarks, like the charged leptons, come in three generations with masses vastly much smaller than the Planck mass. We have found

$$\delta_U^{(1)} = \Delta_U^{(1)} = \frac{\alpha_0}{3\pi}\left(\ln\frac{M_P^2}{m_U^2} + \frac{1}{2}\right) + \frac{3\alpha_0^{(s)}}{4\pi}\left(\frac{4}{9}\right)\left(\ln\frac{M_P^2}{m_U^2} + \frac{1}{2}\right) - \frac{\alpha_0}{\pi}\left[\frac{1}{16}csc^2\theta_W I_1(m_U, m_D) \right.$$
$$\left. + \left(\frac{1}{32}cot^2\theta_W + \frac{17}{288}tan^2\theta_W - \frac{1}{48}\right)I_2(m_U) + \left(\frac{1}{3} - \frac{1}{9}tan^2\theta_W\right)I_3(m_U)\right] \quad (26)$$

for the up-type quarks and

$$\delta_D^{(1)} = \Delta_D^{(1)} = \frac{\alpha_0}{12\pi}\left(\ln\frac{M_P^2}{m_D^2} + \frac{1}{2}\right) + \frac{3\alpha_0^{(s)}}{4\pi}\left(\frac{4}{9}\right)\left(\ln\frac{M_P^2}{m_D^2} + \frac{1}{2}\right) - \frac{\alpha_0}{\pi}\left[\frac{1}{16}csc^2\theta_W I_1(m_D, m_U) \right.$$
$$\left. + \left(\frac{1}{32}cot^2\theta_W + \frac{5}{288}tan^2\theta_W + \frac{1}{48}\right)I_2(m_D) + \left(\frac{1}{6} + \frac{1}{18}tan^2\theta_W\right)I_3(m_D)\right], \quad (27)$$

for the down-type quarks, where the subscripts U and D correspond to up- and down-type quarks, respectively, and where $\alpha_0^{(s)}$ is the bare strong interaction (color) coupling constant. The coefficients of the electroweak contributions in (26) and (27) differ from those in (16) due to different quantum number assignments, and the factor of 4/9 in the strong interaction contributions is a group-theoretical factor that comes from the SU(3) matrices. As in (16), we have neglected the contributions from the Goldstone and Higgs bosons, since the values of the bare quark masses $m_0^{(U)}$ and $m_0^{(D)}$ we have chosen (as discussed below), like that of the charged leptons, correspond to values of the bare quark Yukawa couplings that are much less than the bare electromagnetic and strong coupling constants, and have no numerical significance at this level of approximation.[2]

---

[2] The charged Goldstone bosons give contributions to $m_U$ that are proportional to $m_D$ (and vice versa) which can be accounted for by generalizing (7) to a matrix equation in which $m_U$ and $m_D$ are the elements of a column matrix, but this is unnecessary here since we are neglecting contributions from the bare quark Yukawa couplings.

These expressions have the same structure as (16) and, as in the electroweak theory, generate either one or three physical masses, depending on the values of the bare parameters. We have used the values of the three heavy quark masses $(m_c, m_b, m_t)$, which are known with little ambiguity, to determine values of the two bare quark masses and the bare strong coupling constant:

$$\alpha_0^{(s)} = 0.196; \; m_0^{(U)} = -6.70 \; GeV; \; m_0^{(D)} = 3.45 \; GeV. \tag{28}$$

Due to flavor-changing currents and the presence of the Feynman integral $I_1$, which depends upon both $m_U$ and $m_D$, the self-consistency equations for up- and down-type masses are coupled. For these values of the bare parameters there are three up-type masses for any value of a down-type mass but only one down-type mass for each up-type mass. Consequently, the simultaneous solutions of the coupled self-consistency equations result in three generations of quark doublets. The problem is that when $m_t > M_W + m_b$ the Feynman integral $I_1(m_t, m_b)$ is complex with a small phase ($\approx e^{0.015 \, i \, \pi}$). In this case solutions are complex and there are no simultaneous solutions to the self-consistency equations in which a single up-type mass is coupled to a single down-type mass, as discussed below. For the moment we shall neglect the small complex phase in $I_1(m_t, m_b)$ for simplicity in order to obtain real values of the solutions, in which case each up-type mass is coupled to single down-type mass, as depicted schematically in Figure 3.

Since the weak mixing angle runs with energy, we have had to estimate values of $sin^2\theta_W$ at the masses of the quarks, and from theoretical estimates of the mixing angle in the $\overline{MS}$ renormalization scheme [13] we have used

$$sin^2\theta_W(< 1.5 \; GeV) \approx 0.239; \; sin^2\theta_W(4.5 \; GeV) \approx sin^2\theta_W(173 \; GeV) \approx 0.235. \tag{29}$$

Using the "zero phase" approximation mentioned above, the bare parameters in (28), the mixing angles in (29), and the cutoff $\Lambda = M_P$, the self-consistency equations generate three up-type masses,

$$m_u \approx 309 \; MeV; m_c \approx 1.44 \; GeV; m_t \approx 173 \; GeV, \tag{30}$$

coupled to three down-type masses,

$$m_d \approx 4.0 \; GeV; m_s \approx 4.0 \; GeV; m_b \approx 4.2 \; GeV. \tag{31}$$

If we interpret these physical masses as constituent masses, the solutions for the up-type quarks are a reasonable description of the mass spectrum, although the solutions for $m_d$ and $m_s$ are an order of magnitude too large. However, we have neglected the phase of $I_1(m_U, m_D)$, in which case there is no mixing, so these results are only rough estimates of the true one-loop solutions.

### 7.2 The origin of mixing and the absence of flavor-changing neutral currents

Our approach to the generation problem automatically explains the origin of mixing and the absence of flavor-changing neutral currents. The reason is simple. The bare Lagrangian contains a single bare up-type quark and a single bare down-type quark, each of which is coupled to itself in flavor-conserving neutral currents, and which are coupled to each other in a flavor-changing charged current. There *are* no flavor-changing neutral currents in a theory with only one generation. The effect of mass renormalization is to replace each bare mass by three physical masses without changing the fundamental structure of the theory.

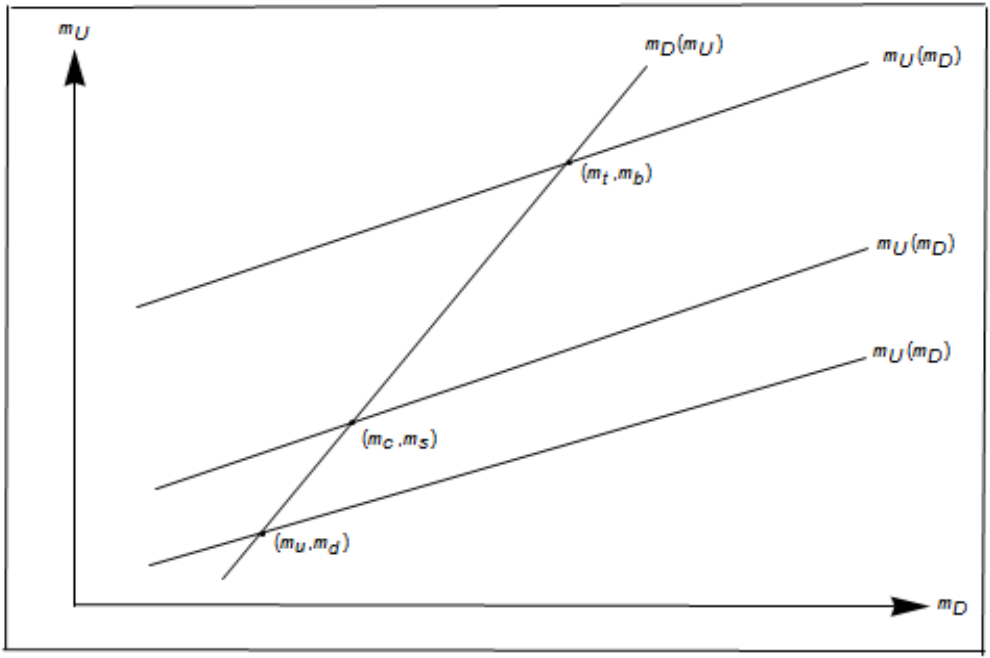

Figure 3. Schematic illustration (not to scale) of the solutions of the one-loop coupled self-consistency equations for quark masses in the $m_U - m_D$ plane using the "zero phase" approximation and the bare parameters discussed in the text. The simultaneous solutions correspond to three generations of quark doublets (six flavors), all of which originate from a universal bare doublet.

This leaves us with three up-type quarks and three down-type quarks, each of which is coupled to itself via flavor-conserving neutral currents, three up-type quarks coupled to three down-type quarks via flavor-changing charged currents, and no flavor-changing neutral currents. The particular combinations of up- and down-type quark states that are coupled to one another (i.e. the mixing angles), like the physical masses themselves, are determined by the multiple solutions of the coupled self-consistency equations, and mixing is required because the solutions are complex.

To see this, consider the basis in which $m_t$ is real, which can be arranged by suitably adjusting the phase of the top quark field. The expression for $\Delta_U^{(1)}$ in (26) only leads to a real solution for $m_t$ if $I_1(m_t, m_b)$ is real, and with $m_t \sim 173\ GeV$ and $|m_b| \sim 4.2 GeV$ there is no phase of $m_b$ for which $I_1(m_t, m_b)$ is real and therefore no solution for $m_t$. The problem can only be avoided by allowing the quarks to mix, so that the expressions for $I_1(m_U, m_D)$ and $I_1(m_D, m_U)$ in (26) and (27) are generalized with the replacements

$$I_1(m_U, m_D) \to \sum_j |V_{ij}|^2 I_1(m_i, m_j) \tag{32}$$

and

$$I_1(m_D, m_U) \to \sum_i |V_{ij}|^2 I_1(m_j, m_i) \tag{33}$$

respectively, where $i = u, c, t$ and $j = d, s, b$, and $V_{ij}$ is the usual CKM matrix that defines the down-type weak eigenstates in terms of mass eigenstates as

$$\begin{pmatrix} |d'\rangle \\ |s'\rangle \\ |b'\rangle \end{pmatrix} = \begin{pmatrix} V_{ud} & V_{us} & V_{ub} \\ V_{cd} & V_{cs} & V_{cb} \\ V_{td} & V_{ts} & V_{tb} \end{pmatrix} \begin{pmatrix} |d\rangle \\ |s\rangle \\ |b\rangle \end{pmatrix}. \tag{34}$$

This leaves us with complex masses $|m_i|e^{i\varphi_i}$ and $|m_j|e^{i\varphi_j}$ and real mixing angles $|V_{ij}|$, whereas the standard model Lagrangian is expressed in terms of real masses and complex mixing angles. However, the phases in the masses can be translated into phases of the mixing angles by absorbing these phases into the left-handed quark fields. If $q_{L_i}$ and $q_{L_j}$ are the left-handed up- and down-type quark fields, one can define new quark fields $q'_{L_i} = e^{-i\varphi_i} q_{L_i}$ and $q'_{L_j} = e^{-i\varphi_j} q_{L_j}$. In terms of the primed quark fields, the mass terms in the Lagrangian are now $\overline{q'_{L_i}}|m_i|q_{R_i}$ and $\overline{q'_{L_j}}|m_j|q_{R_j}$ and the terms in the charged current are $\overline{q'_{L_i}}\gamma_\mu |V_{ij}| e^{-i(\varphi_i - \varphi_j)} q_{L_j}$. The kinetic terms and the flavor-conserving currents are unchanged when expressed in terms of the primed quark fields. Thus, although the solutions of the self-consistency equations involve complex masses and real mixing angles, we can just as well work with real masses and complex mixing angles (which lead to CP violation), as we ordinarily do.

The numerical solutions for the masses and mixing angles will require additional work. However, the self-consistency equations explain the necessity of quark mixing and CP violation as consequences of the standard model dynamics when $m_t > M_W + m_b$.

## Summary

In this paper we have suggested that the parameters of the standard model can be understood dynamically by attributing a physical significance to the ultraviolet divergences. We have done so by treating the Planck mass as a natural ultraviolet cutoff and by rearranging and formally summing the perturbation series for the self-masses of quarks and leptons. The resulting self-consistency equations for the physical masses have enabled us to address the following questions about the standard model:[3]

- Why are there three families of quarks and leptons? ("Who ordered that?")
- Why are they so light? (masses $\lesssim 10^{-17}$ in natural units)
- Why do they exhibit large mass hierarchies? ($m_\mu/m_e \approx 200$, etc.)
- Why do they mix?
- What is the origin of CP violation?
- Why are there no flavor-changing neutral currents?
- What determines the value of the fine structure constant? ($\alpha \approx 1/137$)

---

[3] An extension of the standard model is necessary to account for tiny neutrino masses. However, the techniques developed in this paper could be applied to such an extension.

Within our traditional view of the standard model these are unanswerable questions. At the same time, the divergences are viewed as a fundamental flaw. We have combined these two problems into a solution by using the divergences to develop new relationships between the bare and the physical masses.

Our treatment of the divergences using a Pauli-Villars regulator introduces ambiguities and a lack of gauge invariance in the mass renormalizations of the electroweak theory. We have attributed these to unphysical gauge degrees of freedom that would be absent in a consistent quantum theory of gravity that cuts off the divergences in an unambiguous and gauge invariant fashion. We have chosen to work exclusively in Feynman gauge, which we have argued is equivalent to a physical, gauge-free formulation of the theory. On the other hand, the usual renormalization program, successful though it is for calculating scattering amplitudes in terms of the physical parameters, is tantamount to ignoring the divergences; in the present approach we work directly with the divergences and treat them as large compared to the much smaller cutoff-independent contributions.

Of course, it may be that some underlying substructure of quarks and leptons will eventually explain the parameters in terms of some more fundamental constituents. However, no evidence has emerged so far to suggest anything but a point-like structure of quarks and leptons, which unavoidably leads to divergences. If this is really the case, and the quarks and leptons are truly fundamental point-like objects, the simplest explanation of their parameters appears to be a self-consistent description using the equations that describe their dynamics, divergences and all.